\documentclass[a4paper,12pt]{article}
\pdfoutput=1
\usepackage{graphicx, rotating,amssymb,amsmath,soul}
\usepackage{jcappub}
\include{aastex_journal_defs}

\usepackage{hyperref}

\newcommand{\fref}[1]{Fig.~\ref{#1}}

\begin{document}

\begin{flushleft}
\scriptsize{
LAPTH-038/14}
\end{flushleft}

\vspace{-0.02cm}

\begin{center}

{\Huge\bf Galactic Center gamma-ray ``excess'' from an active past of the Galactic Centre?}

\medskip
\bigskip\color{black}\vspace{0.4cm}

{
{\large\bf Jovana Petrovi\'c}$^{\,a}$,
{\large\bf Pasquale D. Serpico}$^{\,b}$,
{\large\bf Gabrijela Zaharijas}$^{\,c,d}$
}
\\[7mm]

{\it $^a$ {\href{http://www.pmf.uns.ac.rs/en/about_us/departments/physics}{Department of Physics}, Faculty of Sciences, University of Novi Sad, Trg Dositeja Obradovi\'ca 4, 21000 Novi Sad, Serbia}}\\[3mm]
{\it $^b$ Laboratoire de Physique Th{\'e}orique d' Annecy-le-Vieux (\href{http://lapth.cnrs.fr/}{LAPTh}), Univ. de Savoie, CNRS, B.P.110, Annecy-le-Vieux F-74941, France}\\[3mm]
{\it $^c$ Abdus Salam International Centre for Theoretical Physics (\href{http://www.ictp.it}{ICTP}),
	Strada Costiera 11, 34151 Trieste, Italy}\\[3mm]
{\it $^d$ Istituto Nazionale di Fisica Nucleare - Sezione Trieste (\href{http://www.ts.infn.it/}{INFN}),\\
		Padriciano 99, I - 34149 Trieste, Italy}

\end{center}


\bigskip

\centerline{\large\bf Abstract}
\begin{quote}
\color{black}
\large
Several groups have recently claimed evidence for an unaccounted gamma-ray excess over {the} diffuse backgrounds at few GeV in {the} Fermi-LAT data in a region around the Galactic Center, consistent with a dark matter annihilation origin. We demonstrate that the main spectral and angular features of this excess can be reproduced if they are mostly due to inverse Compton emission from high-energy electrons injected in a burst event of $\sim 10^{52}\--10^{53}\,$erg roughly ${\cal O}(10^6)\,$years ago. We consider this example as a proof of principle that time-dependent phenomena need to be understood and accounted for---together with detailed diffuse foregrounds and unaccounted ``steady state'' astrophysical sources---before any robust inference can be made about dark matter signals at the Galactic Center.  {In addition, we point out that the timescale suggested by our study, which controls both the energy cutoff and the angular extension of the signal, intriguingly matches (together with the energy budget) what is indirectly inferred by other evidences suggesting a very active Galactic Center in the past, for instance related to intense star formation and accretion phenomena.}
\end{quote}


\section{Introduction} \label{sec:intro}

The Galactic Center (GC) represents one of the most interesting environments for astroparticle physics studies: it hosts the closest supermassive black hole, which may allow for interesting tests of General Relativity~\cite{2010PhRvD..81f2002M,2011JPhCS.283a2030P}, and it is likely the brightest spot in terms of DM annihilation emission, in models where this
mysterious component is made of weakly interacting massive particles (WIMPs), thermal relics of the Early Universe.
Unfortunately for high energy astroparticle probes, it is also one of the most crowded and hard-to-model regions due to the variety of non-thermal astrophysical sources it hosts,  {thus requiring careful studies before a satisfactory characterization of the signal from this region can be achieved; see for instance the reports from the Fermi-LAT team spanning several years~\cite{Vitale:2009hr,2010arXiv1012.2292M,simona}.}

These difficulties have not discouraged researchers to analyze publicly available Fermi-LAT data\footnote{http://fermi.gsfc.nasa.gov/ssc/data/}, notably looking for these elusive models of physics beyond the standard model. Recently,  several groups \cite{Hooper:2010mq,Boyarsky:2010dr,Hooper:2011ti,Abazajian:2012pn,Hooper:2013rwa,Gordon:2013vta,Macias:2013vya,Daylan:2014rsa,Abazajian:2014fta} found an excess of gamma rays above the modeled astrophysical emission in the inner region of our Galaxy. The claim that it could originate in the annihilation of motivated WIMP scenarios with properties close to commonly expected ones sparked significant attention. The most important properties of the claimed residuals are i) Their spatial extension---which even accounting for the point-spread function is inconsistent with a point-like source---resembles a steeply falling function of the distance $r$ from the GC, behaving as $\sim r^{-2.4}$ and reaching out to  $\sim 10^\circ$ scale. It is also claimed to be close to spherically symmetric, although this statement is probably less robust, due to the difficulty in modeling and subtracting the emission in the Galactic plane. ii) Their spectral shape, which is well modeled by a power law with an exponential cutoff ({\sf PLexp}) 
---of the type $E^{-\Gamma}~\exp\left(-E/E_{\rm cut}\right)$ with parameters in the range $\Gamma =0.5\-- 1$, $E_{\rm cut}\sim 2\-- 3$ GeV---is consistent with the $\sim$0.1 $\--10$ GeV byproducts of a $30-40$ GeV DM particle annihilating dominantly to the $b{\bar b}$ channel. iii) Their total flux at 1-3 GeV, integrated within $1^\circ$ of Galactic Center, is $\sim 10^{-10}$ erg cm$^{-2}$ s$^{-1}$, roughly matching what is expected from a 30 GeV thermal relic DM annihilating with a profile consistent with point~i).

While these findings are quite intriguing, some caveats apply:  all these analyses rely on the publicly available Fermi diffuse model\footnote{http://fermi.gsfc.nasa.gov/ssc/data/access/lat/BackgroundModels.html} to predict the astrophysical signal in that region. While that model is one of the best suited to describe the Milky Way $\gamma$-ray emission, it is obtained via a fit to data for the main purpose of studying {\it point sources} and is therefore not optimal for the characterization of extended signals, which are (at least partially) degenerate with the diffuse emission modeling. Hence the {systematic} errors associated to the separation between signal and background are not yet well assessed.  Even if some extra emission seems to be present, then, some of the above listed properties of the extended residuals {are not robust, at least in their quantitative aspects}. These uncertainties affect even DM interpretations: for example, in~\cite{Cirelli:2013mqa} some of us warned about the importance of the poorly known  bremsstrahlung contribution to the DM signal, especially at GeV and sub-GeV energies, notably for directions along to the Galactic plane. Very recently, in \cite{Gordon:2013vta,Abazajian:2014fta} some effort was put in quantifying some of these systematics, and it was concluded that in particular below 1 GeV the shape of the spectrum is highly model dependent. In the following we will therefore focus especially on the emission at {$E\gtrsim$1-2$\,$GeV. This is also coherent with some other limitations of our model, discussed below.}

On top  of modeling the diffuse astrophysical backgrounds, there is certainly the possibility of competing astrophysical explanations for the residuals. Unresolved milli-second pulsars (MSP) represent one of the most concrete candidates. The Fermi-LAT has turned out to be a {wonderful ``MSP detector'', causing} the discoveries of gamma ray MSP {to skyrocket, also} thanks to the numerous multi-wavelength campaigns~\cite{Caraveo:2013lra}. {The average spectrum of MSPs} is intriguingly similar to the one inferred for the {GC} residuals, being a {\sf PLexp} with $\Gamma \sim 1.5$, $E_{\rm cut}\sim 3.3$ GeV. Again, the situation is far from settled: Based on the number of MSPs which would be observable by Fermi LAT if their unresolved counterparts were to explain the anomalous emission, it was claimed that pulsars {\it can not} make up the excess \cite{Hooper:2013nhl}. However, \cite{Yuan:2014rca} argues that MSPs could make up the diffuse emission, without an inconsistency with the number of the resolved sources, and that the $\propto r^{-2.4}$ decrease of the signal is very similar to the one shown by X-ray binary systems (thought to be related to MSPs). For the emission at the GC, in the inner pixels, it has also been argued that a non-trivial spectrum of the source associated with the SgtA$^{*}$ source is a plausible explanation~\cite{Boyarsky:2010dr}.  

We uncover here one more layer of complexity in the interpretation of the ``excess'': all the above mentioned contributions are by hypothesis {\it steady state}, at least statistically speaking. The high energy sky is however significantly time-dependent, and some nuclear regions in external Galaxies do show major signs of activity. It is natural to attribute at least part of this activity to environmental effects (e.g. accretion on the supermassive black hole, episodes of star formation), rather than intrinsic peculiarities of these objects, thus raising the possibility that the presently quiescent Galactic Center may have been characterized by a violent activity in the past. For most of what follows, we shall keep a very agnostic view of what such an event might have been, just parameterizing it in terms of a few physical inputs, notably: a) a power law injection spectrum of non-thermal electrons. b) the overall energy of this population. c) the look-back time when this burst happened. In order to keep the number of free parameters low, we shall assume some default reference values for the diffusion coefficient and the energy-loss parameter, but will discuss the impact of varying these assumptions. 

{Before engaging in the details of calculation we make a preliminary remark which concerns the goals as well as the intrinsic limitations of our treatment. While we will show that our predictions for the spectra at $E\gtrsim$1-2$\,$GeV are quite robust, this is not the case for the low-energy part of the spectrum. Additionally, since we will be using an effective one-zone model, we are confident to reproduce the key features at intermediate/large angular scales, but the toy-model is likely to fail at small scales (in the  inner degree  around the GC or so) where inhomogeneities in the background radiation and gas density are important and where also the propagation regime might be different (e.g. inhomogeneous diffusion---very different from the Galactic ``average''---or relevant role of convective winds). We stress, however, that the low-energy part and inner pixels  are exactly the ranges where the properties of the residuals are least robustly determined. Thus, the limitations of our treatment actually match the requirements for such a preliminary analysis.}

In Sec.~\ref{formalism} we first introduce our analytical model for the description of the electron spectrum as a function of time, radial distance from the injection point
at the Galactic Center, and energy. The gamma-ray spectrum is then obtained numerically, by convolving this spectrum with the  inverse-Compton (IC) power, which is in
turn an integral of the interstellar radiation field and the IC cross section. Our results are described in Sec.~\ref{sec:results}.  A discussion and our conclusions are in Sec.~\ref{sec:conclusions}, where we also put our findings in the context of other {\it forensic evidences} suggesting a very active Galactic Center over similar timescales as the ones treated here.

\section{Formalism}\label{formalism}

The time-dependent spectrum $Q(E)$ of cosmic ray electrons injected in a bursting episode (a delta function in time and position), propagating via diffusion and $E$-losses in a homogenous medium is well-known from classical literature on cosmic ray astrophysics, see e.g.~\cite{1964ocr..book.....G}. Here, however, we closely  follow reference \cite{1995PhRvD..52.3265A}, which provides a particularly transparent and sufficiently general analytical solution, which will allow us to illustrate the main physical effects analytically. The analytic solution is particularly simple in the regime in which inverse Compton (in the Thomson regime) and synchrotron energy losses are dominant, which for typical choices of interstellar medium parameters is expected to be fulfilled for electrons with energies $\geq$ 10 GeV. As we are interested in $E\simeq 1-10$ GeV gamma rays produced in IC processes and as $E_{IC}\sim (E_e/m_ec^2)^2\, \epsilon_{\rm ISRF}$, our signal will be dominated by the electron population with energies $\geq$ 15 GeV (assuming ISRF photon energies of $ \epsilon_{\rm ISRF}\sim 1$ eV) and therefore this regime should be valid for our purposes (see however \cite{Cirelli:2013mqa} for some limitation to this picture, especially relevant at sub-GeV energies). {We checked explicitly that this expectation is indeed verified, using the equations in~\cite{1995PhRvD..52.3265A} in their general form and accounting for bremsstrahlung energy losses in obtaining the electron spectra. It turns out that for moderate gas densities, of the order of $\leq 1$ cm$^{-3}$, electron fluxes are not affected and the only effect is in yielding additional gamma ray emission correlated with the gas densities. We will comment further on this emission in the Section \ref{sec:results}, when discussing the IC fluxes.}

The energy distribution function of particles at time $t$ post-burst and distance $r$ from the source is given by 
\begin{equation}\label{eq:1}
\frac{dn}{dE_e} (r,t,\gamma)=\frac{N_0~\gamma^{-\alpha}}{\pi^{3/2}~r_{\rm{diff}}^3}\left( 1-b\,t\,\gamma \right) ^{\alpha-2}~e^{ -(r/r_{\rm{diff}})^2}
\end{equation}
where $\gamma =E_e/m_ec^2$ parametrizes the energy $E$ of electrons, $\alpha$ is the index of the electron injection spectrum, $b$ accounts for inverse Compton and synchrotron energy losses as $d\gamma/dt=b\gamma^2$. This parameter can be expressed as $b=5.2\times 10^{-20}~w/\left(1{\rm eV cm^{-3}}\right)~{\rm s}^{-1}$, where the  background energy density $w$ includes the sum of cosmic microwave photons, magnetic field, and starlight/dust contributions as $w=w_{\rm CMB}+w_{\rm B}+w_{\rm OPT}$. Since we are interested in a wider region extending up to $\sim10^\circ$ around the Galactic Center, we choose $w_{\rm OPT}=3$ eV cm$^{-3}$ and $w_{\rm B}=1$ eV cm$^{-3}$ (corresponding to 10 $\mu$ Gauss magnetic fields). These values are higher than the locally measured ones by a factor of $\sim$3  and 2 respectively, but appropriate for the inner 10 degrees of the Galaxy \cite{Porter:2006tb,Crocker:2010xc}.

Now comes an important point: {\it Energy losses} determine the {\it energy cutoff} in the electron spectrum set by the cooling in time $t$ of electrons with formally infinite injection energy as $\gamma_{\rm cut} =(b~t)^{-1}$. 

The {\it spatial extension} of the electron flux (see the exponential term in Eq.~(\ref{eq:1})) is instead determined by the diffusion length $r_{\rm{diff}}$, which is given by

\begin{equation}\label{eq:rdiff}
r_{\rm{diff}} =  2\left( D(\gamma)~t~\frac{1-(1-\gamma/\gamma_{\rm cut})^{1-\delta} }{(1-\delta) \gamma/\gamma_{\rm cut} } \right)^{1/2}\,,
\end{equation}
where the diffusion coefficient $D$ is taken to be $D(\gamma)\simeq D_0~(\gamma/\gamma_*)^\delta$, with $E_*=3$ GeV, $\delta=0.6$ and $D(10$ GeV$)=6\times 10^{28}$ cm$^2/$s. Note that we are interested in electron population with energies $\geq 10$ GeV, for which the power law scaling of the diffusion coefficient is a good approximation. 
As $r_{\rm{diff}}$ explicitly depends on the age of the source $t$, it breaks a degeneracy between the energy loss parameter $b$ and $t$, which determines the spectral cutoff.  We explore this relation among parameters in Section ~\ref{sec:results}. 
The function $r_{\rm{diff}}$ also changes the spectrum in the sense of depleting the low-energy part of $dn/dE_e$ the farther one is from the origin, at a given time,
since less energetic electrons diffuse more slowly.

The total energy output of the source can be found by integrating the source term in volume, time and energy. In our case  of a bursting source which is a delta function in position and time, the volume and time integrals reduce to unity. The energy integral over the {\it injection} spectrum $Q(\gamma)=N_0 \gamma^{-\alpha}$ is
\begin{equation}
E_{\rm tot}=\int _{m_e c^2} ^{\infty} E\,Q(E)dE= N_0\left( m_e c^2\right) ^2 \int _{1} ^{\infty} \gamma ^{1-\alpha}d\gamma
\end{equation}
resulting in the normalization $N_0=E_{\rm tot}~(\alpha-2)/( m_e c^2)^2$.

In order to calculate the Inverse Compton gamma-ray fluxes from this electron population we follow \cite{Colafrancesco:2005ji}. The inverse Compton emissivity (in {cm$^{-3}$s$^{-1}$}) can be written as

\begin{equation}\label{eq:JIC}
J_{IC} (E_\gamma)=\int dE_e~\frac{dn_e}{dE_e}( E_e)~P_{IC}(E_\gamma, E_e)\,,
\end{equation}

where ${dn_e}/{dE_e}$ is given by Eq.~(\ref{eq:1}) and $P_{IC}$ is the inverse Compton power

\begin{equation}\label{eq:PIC}
P_{IC} (E_\gamma, E_e)=c~E_\gamma~\int d\epsilon~n_{\rm ISRF}(\epsilon)~\sigma(E_\gamma, E_e, \epsilon)\,.
\end{equation}

In the above equation, $n_{\rm ISRF}$ represents the density of the Inter Stellar Radiation Field (ISRF), $\epsilon$ the energy of the ISRF photons and $\sigma$ is the differential Klein-Nishina cross section (see \cite{Colafrancesco:2005ji} for more details). Here is an important caveat: in principle, the quantities in Eq.~(\ref{eq:JIC},\ref{eq:PIC}) are dependent on the position,
both via the electron spectrum and via $n_{\rm ISRF}$. However, our solution of Eq.~(\ref{eq:1}) assumes homogeneous properties for propagation, so we use in the following an ``effective'' value for $n_{\rm ISRF}$ (roughly consistent with the used value of $b$), and consider the angular dependence of the IC flux only due to the radial variation of the electron flux.
This approximation is in general better than it appears, as long as one is in a loss-dominated propagation. Lowering $n_{\rm ISRF}$ would enhance the electron flux,
but deplete the emissivity  (and vice versa), with the two effects mostly compensating. This was explicitly seen with the analogous case for bremsstrahlung in~\cite{Cirelli:2013mqa}.  Additionally, we shall illustrate the consequences of variation in the parameters on the different spectra. This approximation is thus sufficient for the purpose of demonstrating the viability of the proposed emission mechanism to explain the data, although no exact match should be expected.

For the spectral density of the ISRF field $n_{\rm ISRF}$ we took the values which correspond to the ones used in the {\sf GALPROP} code\footnote{http://galprop.stanford.edu/code.php}, for the inner Galaxy, as shown in \cite{Porter:2006tb}.

By integrating Eq.~(\ref{eq:JIC}) along the line of sight, we obtain the flux of IC emission at an angular distance $\psi$ from the Galactic plane (which in our spherical symmetric case corresponds to the Galactic latitude, for longitude equal to zero)
\begin{equation}\label{eq:dFdE}
\frac{d\Phi}{dE}_{IC} (E_\gamma, \psi) =\frac{1}{4\pi} \int dl(r,\psi)  J_{IC} (E_\gamma, r)\,.
\end{equation}

We plot fluxes and discuss our results in Section ~\ref{sec:results}, but we can anticipate some qualitative features of the gamma signal. As long as
the quadratic losses happen in the Thomson regime, we expect the IC spectrum to {reflect} the parent electron spectrum. In particular, its cutoff
will be determined by the electron maximal energy $E_{\rm cut}$ {since} $E_{\rm cut,\gamma}~\sim (E_{\rm cut}/m_e)^2\epsilon_{\rm ISFR}$.  For the
application discussed in this article, this is mostly the regime of interest. {For the sake of completeness we remark however that for higher values of the maximal electron energy, $E_{\rm cut} \gtrsim$500 GeV, the IC process  on the bulk of the starlight energy density  takes place in the Klein-Nishina regime. In this case, the spectral shape of the photons would mostly reflect} the background radiation, with its cutoff $E_{\rm cut,\gamma}$ only mildly dependent from $E_{\rm cut}$.

\section{Results}
\label{sec:results}
\begin{figure}[t]
\begin{center}
\includegraphics[width=0.48\textwidth]{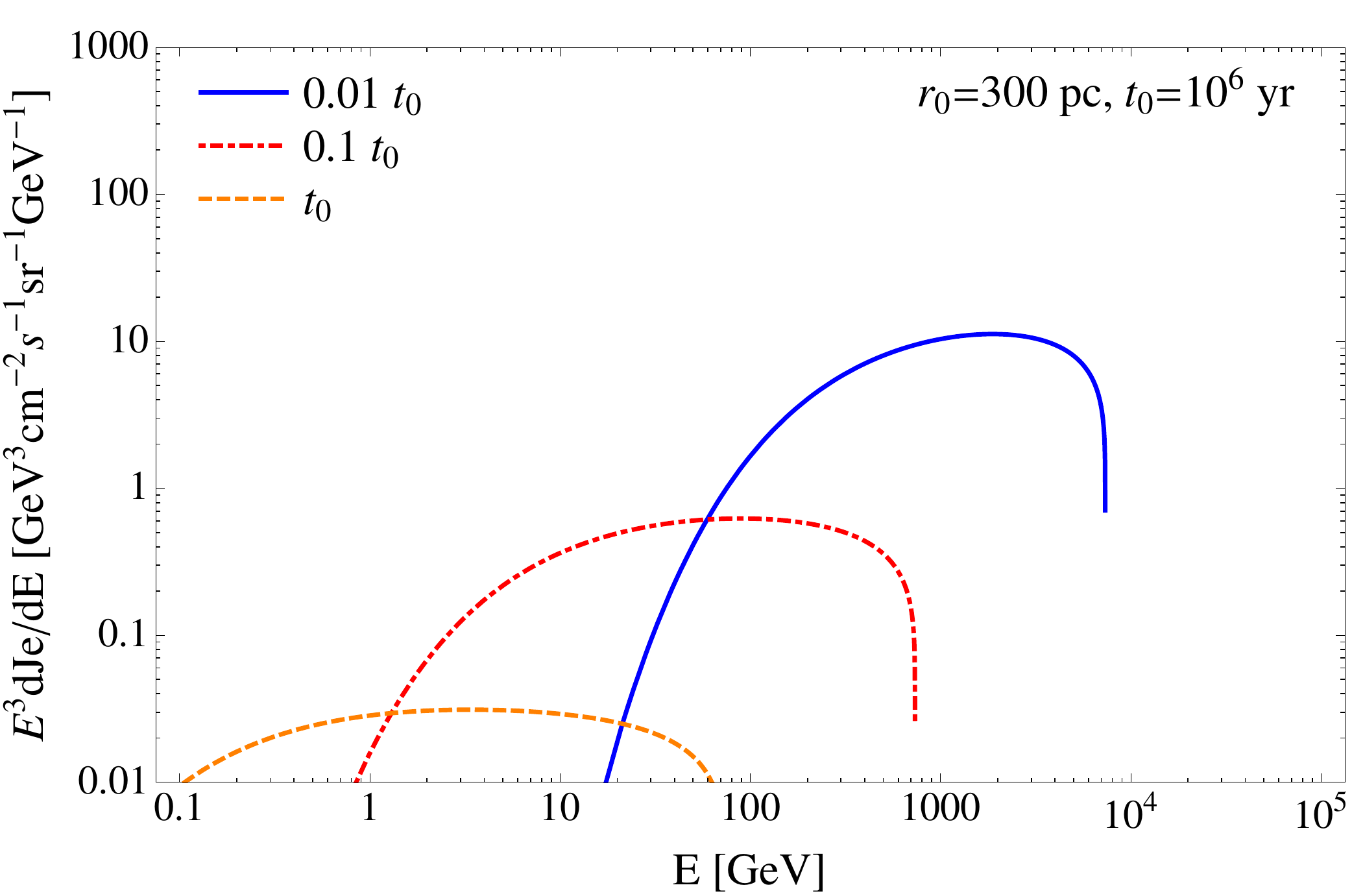}\quad
\includegraphics[width=0.48\textwidth]{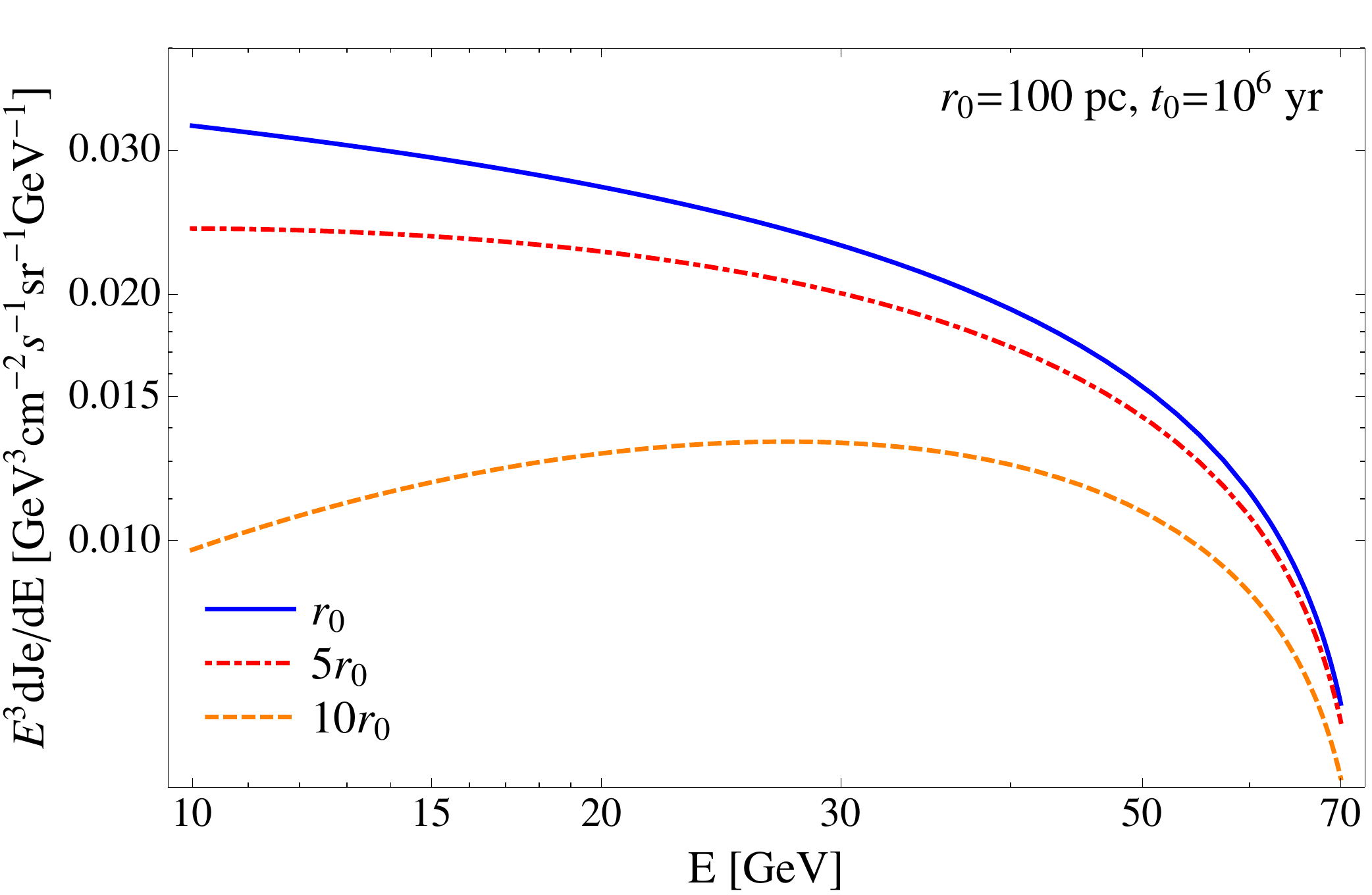}
\caption{Electron spectra from a bursting event at the Galactic Center for different past injection times ({\it {left} panel}) or at different distances from the Galactic Center ({\it {right} panel}). The total energy output is set here to $10^{52}$erg.\label{dndE:varrt}}
\end{center}
\end{figure}

In \fref{dndE:varrt} {\it {left} panel}, we plot the electron fluxes from Eq.~(\ref{eq:1}) for sources at a distance $r_0=100\,$pc, which injected electrons at different times, taken to be $t_0$, 0.1 $t_0$ and 0.01 $t_0$, with $t_0=10^6$yr. The behavior is as expected: the high energy cutoff is set by energy losses, while the fallout at the low energy end is set by the fact that electrons of those energies did not yet reach the distance $r$ to the observer. 
In \fref{dndE:varrt} {\it {right} panel}, we instead keep the injection time fixed and vary the distance to the source. In this case, fluxes farther from the source are lower and especially depleted at low energies, but the energy cutoff stays the same.  

\begin{figure}[!ht]
\begin{center}
\includegraphics[width=0.48\textwidth]{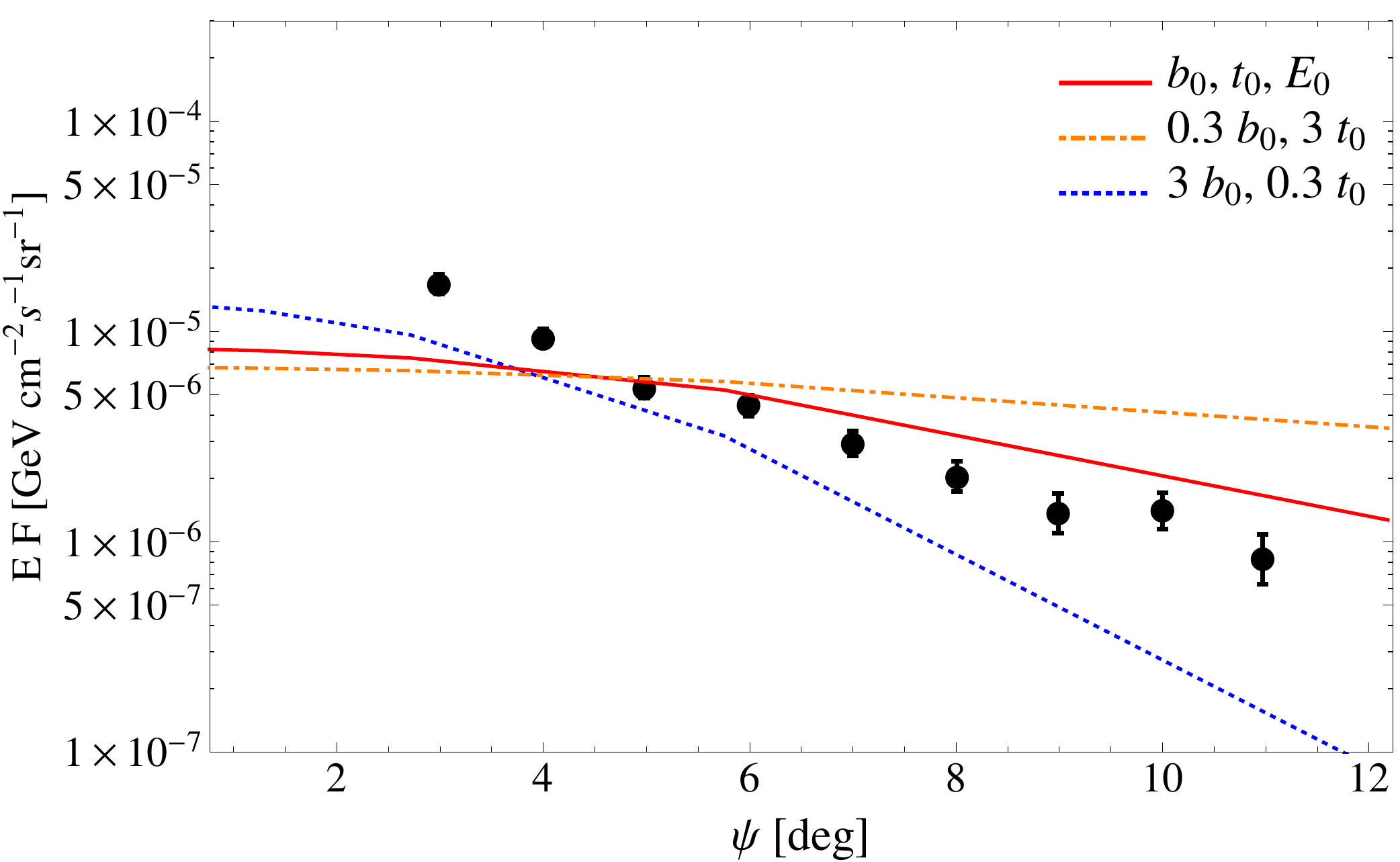}\quad
\includegraphics[width=0.48\textwidth]{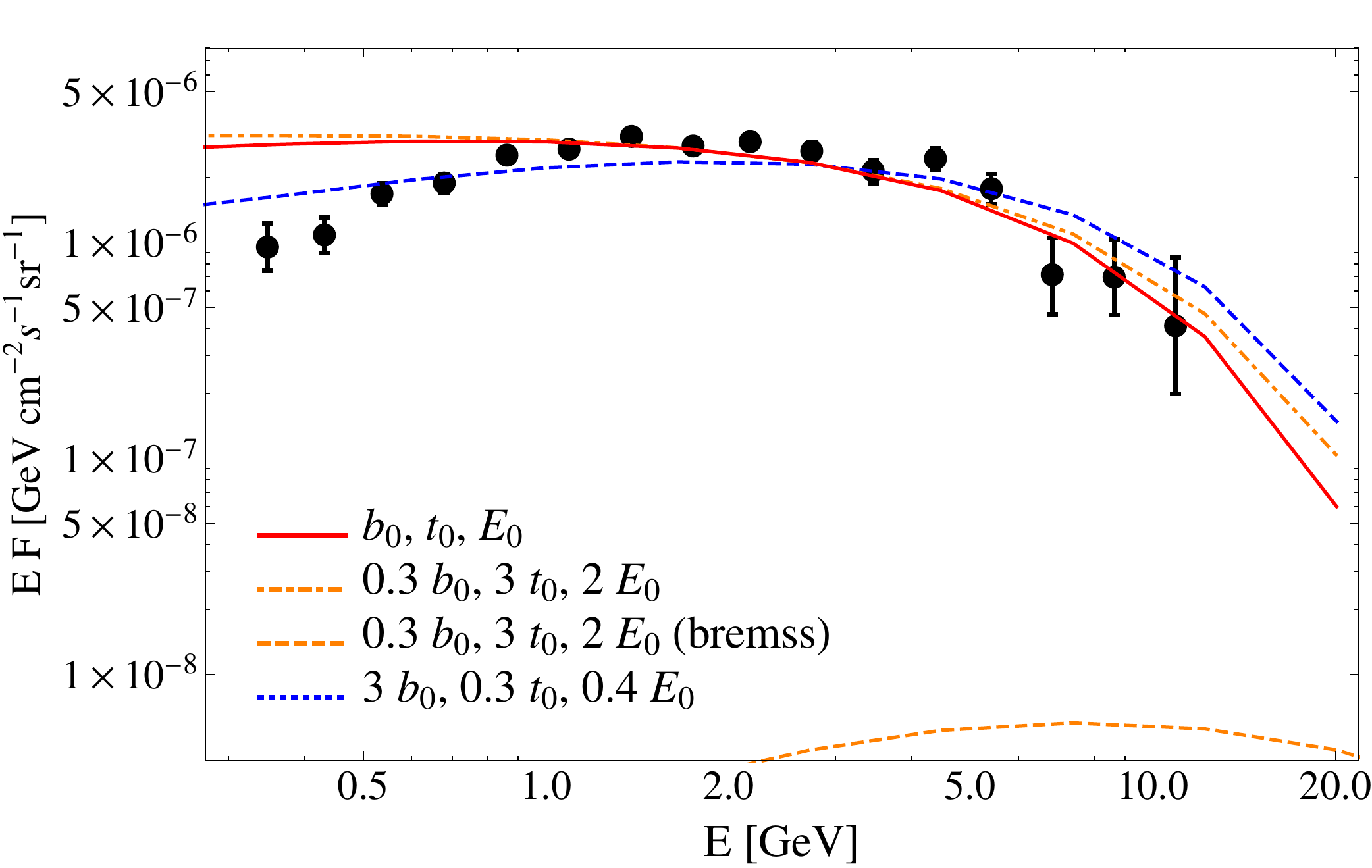}
\caption{{\it {Left} Panel:} Latitude profile of the inverse Compton emission from an electron population injected $t_0$ ({\it red, solid}), $0.3~t_0$ ({\it orange, dashed}) and $3~t_0$ ({\it blue, dotted}) years ago (where $t_0=1$ Myr). {\it {Right} Panel:} The spectra of the inverse Compton emission (the same color scheme) at $5^\circ$ away from the Galactic plane. The overall energetics is given in units of $E_0={3}\times 10^{52}$ erg, and energy losses are expressed in terms of the default value $b_0$, { which assumes $w\sim4$ eV cm$^{-3}$}. {The orange dashed line at the bottom indicates the bremsstrahlung contribution to gamma ray emission $5^\circ$ away from the GC.}  \label{dFdE:varrt}}
\end{center}
\end{figure}

In \fref{dFdE:varrt}  we show the latitude profile calculated at 1 GeV ({\it {left} panel}) and the spectrum 5$^\circ$ away from the Galactic plane ({\it {right} panel}) of the `anomalous' Galactic center emission, as derived in \cite{Daylan:2014rsa} in the fit of a dark matter annihilation template, for which dark matter halo density profile $\sim r^{-1.2}$ was assumed\footnote{Dark matter induced gamma ray fluxes are proportional to the square of the DM density, and therefore scale as $\sim r^{-2.4}$ in this case.}. The error-bars shown are statistical, {and some angular averaging was performed in~\cite{Daylan:2014rsa}, so that the comparison is only indicative}. 
 Note that there is a factor of $\sim 2$ mismatch in the flux normalization at 5$^\circ$ for the fluxes in the two panels. That is a consequence of the fact that the derived flux of the residual emission depends on the assumed morphology of the residuals fitted to the data. We view it as a systematic uncertainty in the fitting procedure which ultimately translates to the uncertainty on the total energetics of the source term. When we quote energetics we will rely on the spectral fluxes in the bottom panel. %

If we keep the spectral injection and diffusion indices fixed to their fiducial values, in principle one has three observables (spectral shape, angular shape and normalization)
with four major parameters ($D_0, ~b,~E_{\rm tot},~t_0$), {of which however only the latter two are related to our source model}. The angular shape is controlled by $r_{\rm{diff}}$, which in turn can be altered via $(D_0\,,t_0)$, see Eq.~(\ref{eq:rdiff}). This singles  out a Myr timescale for $t_0$, for the fiducial value of $D_0$. Once this parameter is fixed, only minor spectral slope adjustments are possible by varying $\alpha \in$ $[2.1\,, 2.4]$, and $D_0(4~{\rm GV})\in [2\times 10^{28}\,,10^{29}]$cm$^2$s$^{-1}$~\cite{FermiLAT:2012aa}, but the key spectral parameter, the cutoff energy $E_{\rm cut}=m_e(b\, t_0)^{-1}$, is determined by {the same parameter}, $t_0$. It is remarkable, we believe, that the observed cutoff in the spectrum is fully consistent with this estimate, as shown in the bottom panel of \fref{dFdE:varrt}.

\begin{figure}[!ht]
\begin{center}
\includegraphics[width=0.48\textwidth]{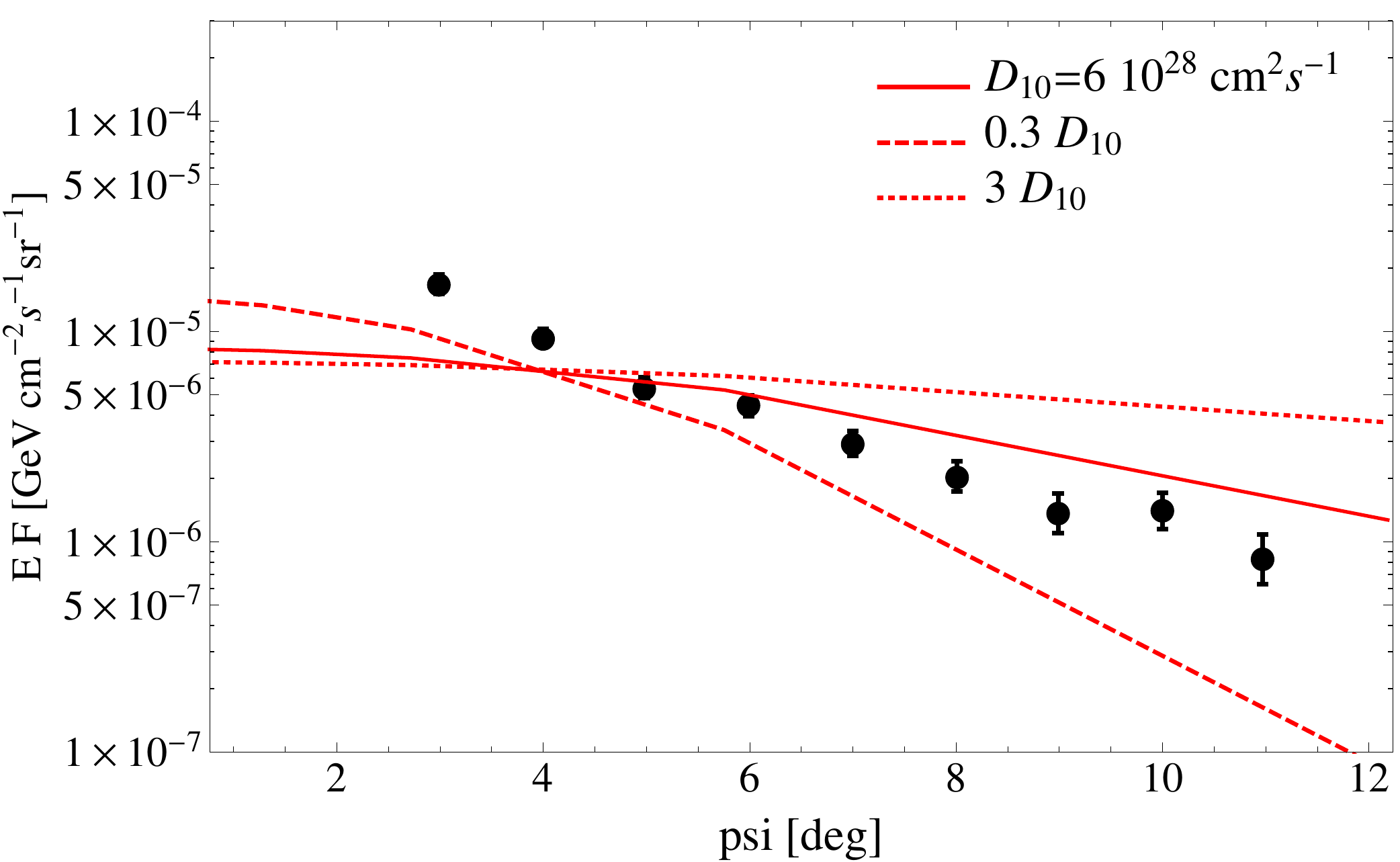}\quad
\includegraphics[width=0.48\textwidth]{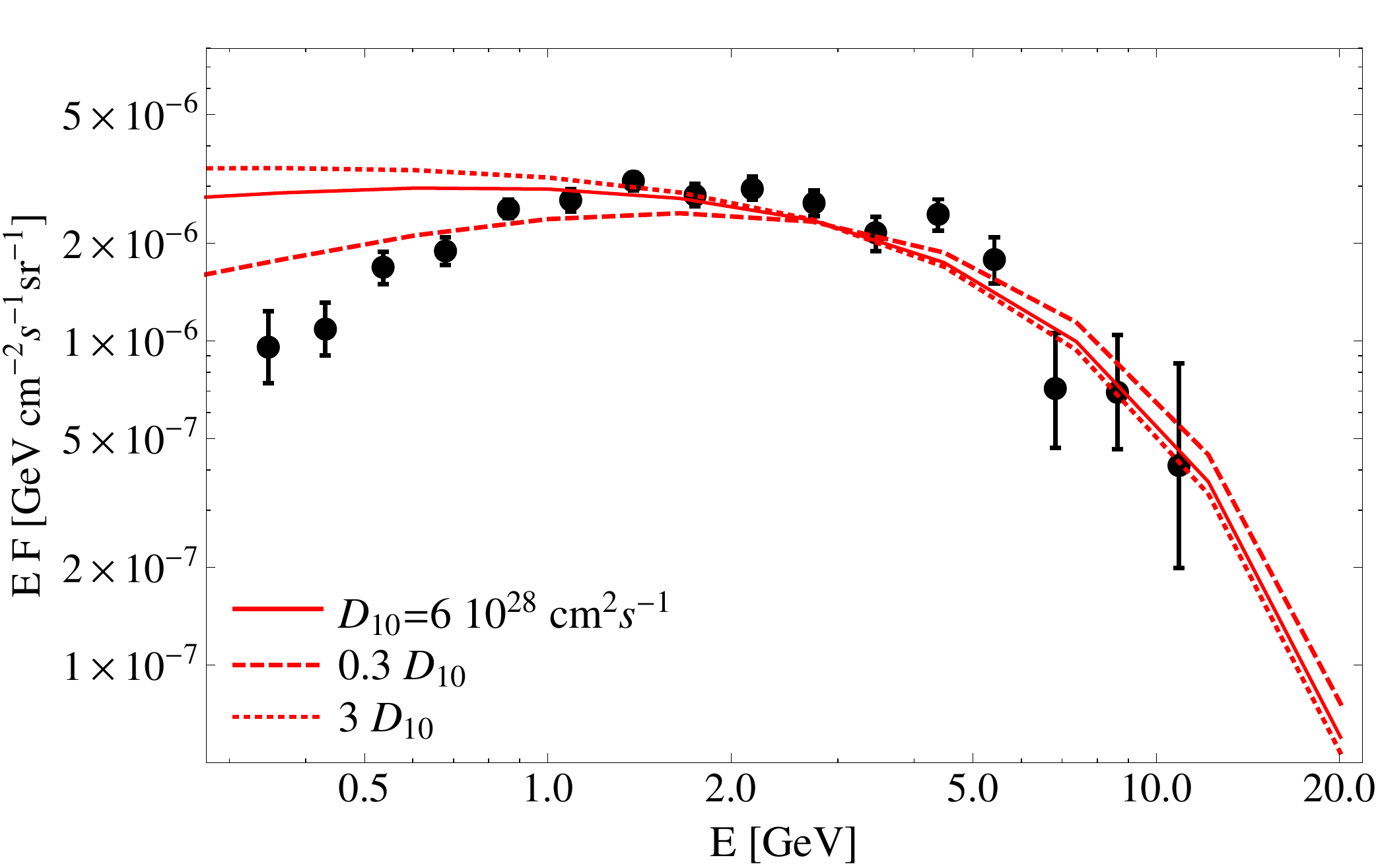}
\caption{Latitude profile ({\it {left}})  and the spectra of the inverse Compton emission at $5^\circ$ away from the Galactic plane ({\it {right}}), for the electron population injected $t_0=1$ Myr ago, with a source of $E_0={3}\times 10^{52}$ erg, calculated with our default values for the set of parameters ({\it solid}). In addition, the diffusion index is varied to $0.3~D_0$ ({\it dashed}) and $3~D_0$ ({\it dotted}), where $D_0 ~(10~{\rm GeV})=6\times 10^{28}$ cm$^2$s$^{-1}$. \label{dFdE:checksD}}
\end{center}
\end{figure}

\begin{figure}[!ht]
\begin{center}
\includegraphics[width=0.48\textwidth]{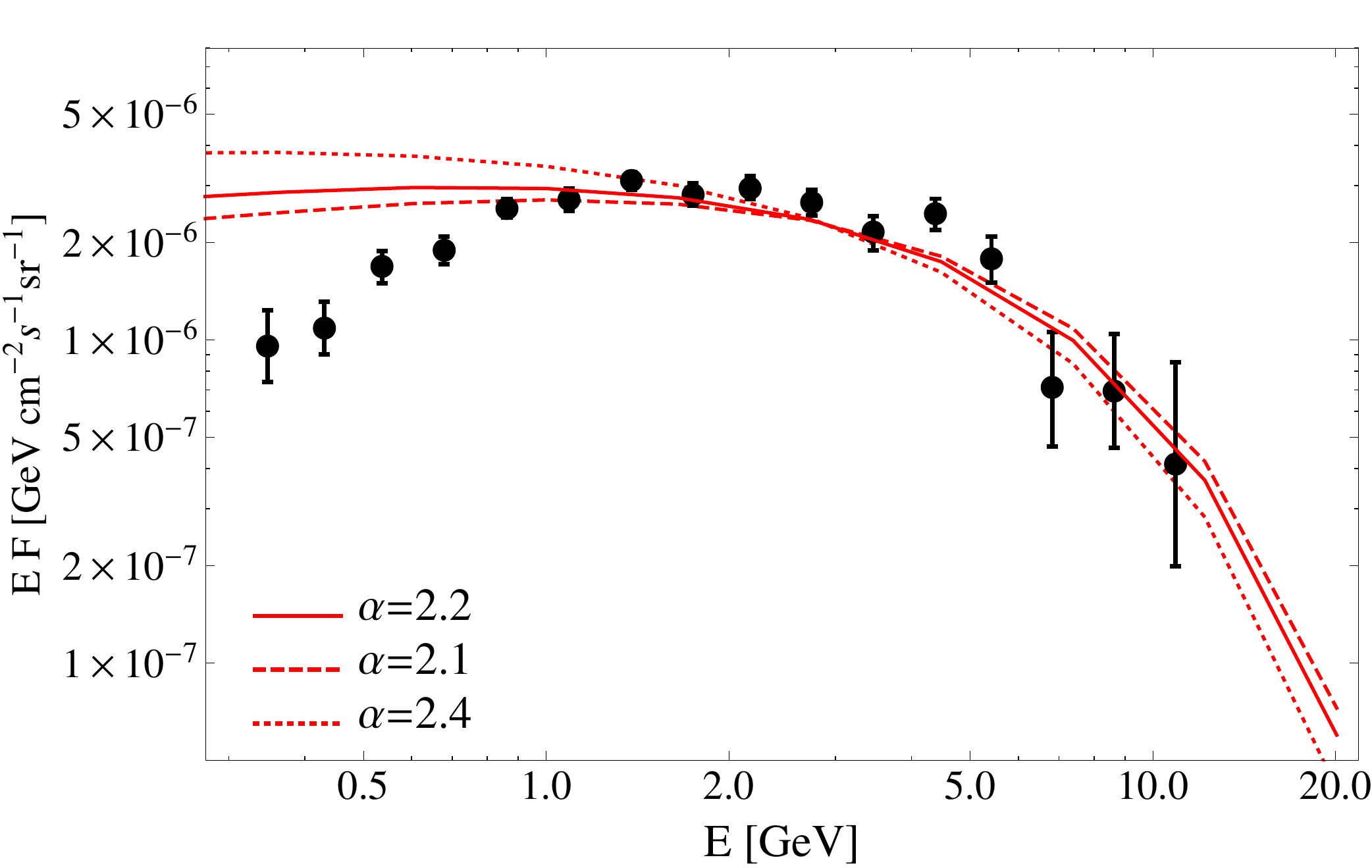}\quad
\caption{{\it Solid line:} The spectra of the inverse Compton emission at $5^\circ$ away from the Galactic plane, for the electron population injected $t_0=1$ Myr ago, with a source of $E_0={3}\times 10^{52}$ erg, calculated with our default values for the set of parameters. The spectral injection index is varied to $\alpha=2.1$ ({\it dashed}) and $\alpha=2.4$ ({\it dotted}). \label{dFdE:checksalpha}}
\end{center}
\end{figure}

In Figs.~\ref{dFdE:checksD} and \ref{dFdE:checksalpha} we instead show how the results change if $\alpha$ and $D_0$ parameters are varied, while the rest of parameters are kept fixed to their fiducial values. { It can be seen in \fref{dFdE:checksD} that the value of the parameter $D_0$ has a larger impact on the latitude profile than on
 the energy spectrum, as expected. This exercise also suggests that there is at least an uncertainty of a factor $\sim3$   on the age of the bursting event around the benchmark value $t_0$=Myr, since the two parameters are almost degenerate.}

{We also calculated the bremsstrahlung gamma-ray emission from this population of electrons, following \cite{Cirelli:2013mqa} and adopting gas densities as implemented in the {\sf GALPROP} code (see Fig. 2 of \cite{Cirelli:2013mqa}). The densities of molecular and atomic gas fall off  rapidly at a vertical distance $\sim 200$ pc$\sim  {\cal O}(1^\circ)$ off the Milky Way plane. At higher latitudes some ionized hydrogen HII is nonetheless present, but with densities more than an order of magnitude lower than those of the other two components. As a consequence, the bremsstrahlung emission at $5^\circ$ away from the plane only amounts to$\sim \cal{O}$(1 $\%)$ of the Inverse Compton flux, as shown in a lower panel of \fref{dFdE:varrt}. It is worth noting that the bremsstrahlung emission does not share the same energy cutoff as the IC emission, extending to energies much closer to the cutoff in the electron spectrum, which is at $E\geq$ 50 GeV in the cases considered here. For choices of environmental parameters different from the benchmark used here, or in the inner $\sim 1^\circ$, this may lead to an observable effect.}

{In summary, we showed that a simplified, homogeneous one zone model of a single bursting source manages to explain several features of the claimed excess, notably
the medium-large scale extent of the emission and the high-energy part of the bump, covering the turning point and  the cutoff.
As argued previously, the model has its own limitations and is expected to be much less predictive in the inner degrees or at energies  below 1 GeV, where however contributions from unresolved point-sources and the finite angular-resolution of Fermi-LAT make the determination of the excess quantitatively much harder in the first place. 
At the same time, while the model sketched here is not very predictive in those ranges,  the shown dependences of the solution on some key parameters suggest  that natural extensions of the model {\it could} accommodate for the (tentative) observations: a inhomogeneous diffusion coefficient, with smaller values in the inner region, could account for the very central excess {\it and} the harder spectrum below the GeV. The radial-dependence of the energy-loss coefficient could also contribute in a similar way. Again, it is non-trivial that 
varying the parameters in a physically realistic direction seems to improve at the same time both the angular and spectral agreement of the predictions with the data.  
Introducing several bursting events with different timescales could also play a role. Finally, we also checked that replacing the featureless power-laws slightly softer than $E^{-2}$---inspired by the non-relativistic first order Fermi acceleration model---considered here with more flexible injection spectra, notably of the form $E^{-\beta}~\exp\left(-E/E_{*}\right)$ with a harder power-law index like $\beta\sim 1.5$ can also improve the spectral agreement at low energies, at the expenses of adding an explicit injection cutoff parameter $E_{*}\sim {\cal O}(100)\,$GeV.}

Note that, depending on the acceleration mechanism (leptonic or hadronic) a bursting event could  also inject a population of high energy protons in the medium, which would as well produce gamma ray emission and additional secondary electrons in the interactions with the interstellar gas. In that scenario, the considerations developed here should be modified, notably because of  the much longer energy loss timescales (proton propagation is typically diffusion dominated) and because of the different efficiency in generating gamma-ray radiation. Additionally,  gamma ray emission would correlate with the gas distribution, which is not the case for the model here. In this article we do not consider a hadronic scenario further, but it is plausible that it could contribute as well (or alternatively) to similar phenomenology~\footnote{{\it Note added:} While this work was being finalized for submission, an in-depth study of this effect has appeared as pre-print~\cite{Carlson:2014cwa}.}. In the recent analysis~\cite{Yoast-Hull:2014cra} -- which provides yet another
argument in favor of the existence of an additional cosmic-ray population in the inner Galaxy---a leptonic scenario for
the underlying population was also considered more likely, based on energetics.

\section{Conclusions}
\label{sec:conclusions}
In this article, we have argued that a bursting event, injecting $\sim10^{52}\,\-- 10^{53}$ergs of energy in a standard power-law cosmic ray electron spectrum about one million years
ago seems to reproduce naturally most spectral and angular features of the claimed GeV ``excess'' in the inner Galaxy, for benchmark values of an effective
homogeneous diffusion coefficient and energy loss parameter.  The main goal of our calculations has been to raise awareness on the importance of accounting for transient events when dealing with extended excesses, notably at the GC. Until now, however, we have not discussed the plausibility of the parameters required. 
Is the ``toy solution'' found plausible, on the light of other astrophysical evidence? After all, currently the GC is best characterized by the quiescent state of its supermassive black hole, see e.g.~\cite{Genzel:2010zy}. Intriguingly, however, there are many hints that the GC may have experienced an active past{, see \cite{Ponti:2012pn} for an overview of the topic}. 

For example, it appears established that two OB stellar clusters formed in the inner parsecs $6\pm 2$ Myr ago, probably triggered by some dense gas accretion event~\cite{2006ApJ...643.1011P}.  Similarly, it has been argued that the degree-scale
Galactic center lobe (GCL) may be associated with the past release of few times 10$^{52}$\,erg burst event according to the multi-wavelength analysis in~\cite{2010ApJ...708..474L}.
Further, there are a number of observables at different scales (arc-minute to tens of degrees) suggesting activity periods in the inner Galaxy, see~\cite{2006PASJ...58..965T} for a review. 
It has even been proposed that ``GC shells'' are related to a {\it Recurrent Starburst Model}~\cite{Sofue:2003xm} where explosive events ranging in timescale from $10^5$ years to $>10^7$ years, and energetics from $10^{51}$ ergs to $10^{55}$ erg, may account for a variety of structures and observations. Perhaps the most spectacular manifestation of an activity in the inner Galaxy is provided by the recently discovered ``Fermi-LAT bubbles''.  Ref.~\cite{Su:2010qj} contains a description of these features as well as a nice review
of different evidence for variability at the GC. Note that the energy budget and the morphology of the GC excess are much less demanding than Fermi-LAT bubbles, and seem
to fall nicely in the ball-park of what expected from a number of dynamical models for the activity at the GC. Needless to say, the model described here is  a simplification: for example residuals could be  due to {\it repeated} bursting events, with different energetics; younger ones would give steeper profile in the GC and be less extended, while older ones would be responsible for the extended emission. Another possibility is to have a combination of unresolved point sources with peculiar spectra, especially near the GC, plus extended emission. Inhomogeneities in the energy losses and diffusion coefficient can also alter to some extent morphology and spectral shape of the signal.
Disentangling among different possibilities might be challenging with limited information, yet {there appears to be no intrinsic physical limitation to such a kind of diagnostics. 
For instance, hadronic models should show some correlation with the gas distribution, since this provides the target for pion production. To differentiate hadronic models from the scenario considered here, it would be very helpful to focus on the highest energy part of the excess, where the angular resolution of the detector is better. The radio signal, which at the largest scales is associated to any form of leptonic emission, could provide another handle:
it has been noticed that radio emission may be very constraining for DM models, in particular if one focuses on the very central regions of the Galaxy~\cite{Bringmann:2014lpa}. 
In our scenario these bounds are naturally evaded, since DM models are steady state and are associated to a GC ``spike'', while in a time dependent picture the densities of leptons in the central region well after the burst events are much more modest.}
Once the existence and characteristics of an extended GeV component will be established more robustly, extensive numerical simulations of the expected signals would be opportune. {We expect that separating this scenario from some unresolved component as MSP will be challenging, and probably will depend on the achievement of a better understanding of the population properties of these objects. A peculiar signature which might arise in the scenario considered here, however, is a 
a departure from a featureless, smooth decline of the excess with distance from the GC, contrarily to DM models, or even to astrophysical steady state models where the injecting population has some monotonically decreasing source density. This could be due to the superposition of several bursting events at different epochs and with different energetics. Provided that the spectrum and the angular distribution of the excess can be measured with sufficient accuracy, one may expect ``shoulders'' and ``bumps'' to start being visible. 
The mechanism discussed here also shares the following features with any other relying on {\it secondary} production from {\it electrons}: i) some degree of non-universality in the spectrum, although likely not sufficient to be resolved clearly, yet. For instance, if the spectra are matching exactly at 5 GeV, the difference between the spectrum integrated over a 1-10$^\circ$ region and the spectrum integrated over a  1-5$^\circ$ region around 1 GeV amounts to $\leq $10\%. These differences would naively grow by a factor of few at lower energies or if one were to compare the inner degree spectrum with the larger scale one, but unfortunately these are the regimes where our scenario is less predictive as explained above, and while the trends may be correct, the quantitative estimate is less reliable. ii) provided that the underlying electron population extends to sufficiently high energies, the presence of some residual emission beyond the cutoff of the IC should be present, due to bremsstrahlung onto the gas. This is barely visible in the right plot of \fref{dFdE:varrt}, but should be more prominent and possibly detectable in the inner regions, where a larger gas density is present, or if current baseline gas maps underestimate the actual HII gas density.  In DM models we expect the features i) and ii) to be much less prominent, at least at sufficiently high energy, due to the large component of prompt gamma-rays and the intrinsic cutoff given by the DM mass. }

What appears certain is that a minimal relaxation of the theoretical modeling assumptions  (i.e. dropping the stationarity assumption) is sufficient to bring new realistic alternatives to the DM hypothesis in fitting the excess at the GC. {Needless to say, the most effective method to check} the DM interpretation {would be to collect}  independent evidence from other observables, including direct detection at underground detectors and collider searches.  {For the time being, we tentatively interpret the lack of counterpart to the putative DM excess in any other charged cosmic ray signal, such as positrons and especially antiprotons~\cite{Bringmann:2014lpa,Cirelli:2014lwa}, as indirectly supporting  astrophysical explanations, which should thus be explored further.}

{\small {\it Acknowledgements:} We are grateful to Seth Digel, Dan Hooper, Tim Linden, Tijana Prodanovi\'c and Andy Strong for instructive comments and to Stefano Profumo for notifying us of their work~\cite{Carlson:2014cwa}. {We are also thankful to Francesca Calore and Veerle Tammer for pointing out a minor error in the results displayed in Fig. 1.} JP is grateful to the ICTP for hospitality during completion of this project.



\end{document}